\documentclass[prd,aps,a4paper,superscriptaddress,twocolumn,nofootinbib]{revtex4}
\usepackage{graphicx}
\usepackage{color}
\usepackage{dcolumn}
\usepackage{bm}
\usepackage{slashed}
\usepackage{amsmath}
\usepackage{latexsym}
\usepackage{amssymb}
\usepackage{mathrsfs}
\usepackage{amsfonts}
\usepackage{url}
\allowdisplaybreaks
\begin{document}

\title{Dark matter minispike: a significant enhancement of eccentricity for intermediate-mass-ratio-inspirals}
\author{Xiao-Jun Yue}
\email[Xiao-Jun Yue: ]{yuexiaojun@tyut.edu.cn}
\affiliation{College of Information and Computer, Taiyuan University of Technology, Taiyuan 030024, China}
\author{Zhoujian Cao
\footnote{corresponding author}} \email[Zhoujian Cao: ]{zjcao@amt.ac.cn}
\affiliation{{Department of Astronomy, Beijing Normal University,
Beijing 100875, China}}
\date{\today}

\begin{abstract}
When a stellar massive compact object, such as a black hole (BH), inspirals into an intermediate massive black hole (IMBH), an intermediate-mass-ratio-inspiral (IMRI) system forms.  Such kind of systems are important sources for space-based gravitational wave detectors including LISA, Taiji and Tianqin. Dark matter (DM) minispikes may form around IMBHs. We study the effect of dynamical friction against DM minispike on the evolution of eccentric IMRI. For such investigation we construct the dynamical equations which describes the evolution of eccentric IMRI under the effect of dynamical friction. As dynamical friction is large for small velocity, the dissipation of energy near apastron is more than that near periastron. This will greatly enhance the eccentricity. For instance, with an initial semi-latus rectum of $1\rm AU$, even a moderate DM minispike can make the eccentricity grow from $0.3$ to $0.85$. In the extremal case the eccentricity could be enhanced to near $1$. We also study a specific case which corresponds to an IMRI in the center of a globular cluster (GC) and find the eccentricity can keep its value above $0.95$ until the IMRI enters LISA band. These gravitational wave with enhanced eccentricity by DM minispikes can be easly distinguished from that without DM at $10^{-3}\rm Hz$ due to the eccentricity difference. These anticipations can be tested by future space-based GW detectors such as LISA.
\end{abstract}

\pacs{98.80.-k,98.80.Cq,98.80.Qc}

\maketitle

\section{\label{s1}Introduction}
A large number of astrophysical and cosmological observations provide convincing pieces of evidence for the existence of dark matter (DM), and the distribution of DM is a subject of great interest. Navarro, Frenk and White obtained a universal density profile for DM halos via cosmological N-body numerical simulations called NFW profile \cite{Navarro_1997}. Gondolo and Silk suggest that if a massive black hole (MBH) exists at the center of the galaxy, the adiabatic growth would modify the DM distribution around the MBH, leading to a density cusp called DM spike \cite{PhysRevLett.83.1719}. It was subsequently shown that dynamical processes like off-center formation of the seed black hole, major merger events of the host galaxies and  gravitational scattering of stars may lead to destruction or reduction of the DM spikes \cite{PhysRevD.64.043504,PhysRevLett.88.191301,PhysRevLett.92.201304,PhysRevD.72.103502}. Even so, the DM ``minispikes'' surrounding the intermediate massive black holes (IMBHs) with mass between $10^2M_\odot$ and $10^6M_\odot$, may survive as major mergers are less likely to destroy the DM cusp around the IMBH.

The discovery of GWs provide us with a new method to observe the Universe. Recently the possibility of probing DM with GW experiments has been under discussion. If the mass distribution of DM could have an influence on the orbits of stars and other matter, it may leave a sign in GWs. In \cite{PhysRevD.89.104059} it was concluded that a survey of the environmental corrections including DM halos to the GW signal are typically negligible for most LISA sources. But in \cite{PhysRevLett.110.221101} it was shown that the gravitational potential of DM minispike around IMBHs can really impact the GW waveform, and the deviation is detectable by LISA. Furthermore, in \cite{PhysRevD.91.044045} it was also suggested that the effect of dynamical friction of DM minispike can also impact the waveform. In \cite{PhysRevD.97.064003} the combined effect of gravitational pulling, dynamical friction and accretion of DM particles was considered, and it was found that the dynamical friction predominates. The results in \cite{Yue_2018} indicate that the merger rate of intermediate-mass-ratio-inspirals (IMRIs) can be increased by DM minispikes, which may affect the event rate of LISA. In \cite{PhysRevD.98.023536} the influence of DM ``dress" on the merger rate of primordial BHs are also investigated.

The DM minispikes may have an influence on the GW waveforms, but the works in \cite{PhysRevLett.110.221101,PhysRevD.91.044045,PhysRevD.97.064003} only investigated quasicircular orbits. It is expected that gravitational waves from IMRIs moving in highly eccentric orbits are excellent sources for LISA \cite{Gultekin_2006,Matsubayashi_2007,G_rkan_2006}. To detect GWs, one needs to know a priori of the binary's orbital evolution which strongly affects the inspiral waveform, but the dynamical equations for eccentric IMRIs affected by DM minispikes are still unestablished.

In the current paper we investigate the dynamics of an eccentric IMRI affected by a DM minispike in the Newtonian formalism. We focus on the dynamical friction effect as it is dominant \cite{PhysRevD.97.064003}. Based on the adiabatic approximation, we derive the dynamical equations for an eccentric IMRI affected by a DM minispike. As the dynamical friction is inversely proportional to the square of velocity, the effect is stronger for larger radius. This makes the DM minispike tend to increase the eccentricity when the IMRI inspirals in. In the evolution history of the binary, the eccentricity can be increased to approach to $1$ in some extremal cases. We also study a specific case of an IMRI in a typical globular cluster (GC). We find that the DM minispike can increase the eccentricity  up to $0.95$ at the gravitational wave frequency $10^{-4}\rm Hz$, which is the lower side of LISA band. Such a large eccentricity will result in a series of cusps in GW waveforms, which can be used to distinguish the existence of DM minispike.

This paper is organized as following. In the Sec.~\ref{s2} we will derive the dynamical equations for an eccentric IMRI affected by a DM minispike. In the Sec.~\ref{s3} we will study an specific IMRI in a typical globular cluster (GC) and investigate its related GW waveform in LISA band. The Sec.~\ref{s4} is for summary and conclusion.

\section{\label{s2}Dynamical Equations for eccentric IMRI}

We consider a DM minispike model used in \cite{PhysRevD.91.044045}.  The DM minispike has a spherically symmetric distribution with a single power law
\begin{equation}
\rho_{\rm DM}(r)=\rho_{\rm sp}\left(\frac{r_{\rm sp}}{r}\right)^\alpha.\label{eq1}
\end{equation}
The $r_{\rm sp}$ is the radius of the DM minispike and $r_{\rm sp}\approx 0.2r_h$ where  $r_h$ is the influence radius of the IMBH with the relation $M(<r_h)=\int_0^{r_h} 4\pi\rho(r)r^2 dr=2M$, with $M$ being the mass of the IMBH. $\rho_{\rm sp}$ is the DM density at the radius $r_{\rm sp}$. Following the setting used in \cite{PhysRevD.91.044045}, we adopt $r_{\rm sp}=0.54\rm pc$ and $\rho_{\rm sp}=226M_\odot/\rm pc^3$.

The power law index $\alpha$ of the minispike depends on the initial DM halo prior to the formation of the IMBH. If the initial DM halo has an NFW profile with power law index $\alpha_{\rm ini}=1$, after the adiabatic growth of the IMBH the parameter $\alpha=(9-2\alpha_{\rm ini})/(4-2\alpha_{\rm ini})=7/3$ \cite{Navarro_1997}. If the initial halo has a uniform density distribution, $\alpha=1.5$ \cite{PhysRevD.64.043504,1995ApJ...440..554Q}. In this paper we assume $1.5<\alpha<7/3$.

We consider a binary system which involves a small BH with mass $\mu=10M_\odot$ orbiting around an IMBH with mass of $M=10^3M_\odot$. As the mass of the secondary object is much smaller than the IMBH, the reduced mass is approximately $\mu$ and the total mass is approximately $M$. We can also take the barycenter to be approximately at the position of the central IMBH.

Based on the Newtonian mechanics, the orbit of the small BH lies in a plane, and the orbital angular momentum is
\begin{equation}
L=\mu r^2\dot{\phi},\label{eq2}
\end{equation}
where $r$ is the seperation of the binary and $\phi$ is the angular position of the small BH. The total energy is given by
\begin{eqnarray}
E=\frac{1}{2}\mu(\dot{r}^2+r^2\dot{\phi}^2)-\frac{G\mu M}{r}\nonumber\\
=\frac{1}{2}\mu\dot{r}^2+\frac{L^2}{2\mu r^2}-\frac{G\mu M}{r}.\label{eq3}
\end{eqnarray}
In the second step we have used the Eq.~(\ref{eq2}). Without the consideration of dissipative processes, the energy $E$ and angular momentum $L$ are conserved.

Any bounded equatorial orbit can be described by the semi-latus rectum $p$ and the eccentricity $e$. These two quantities are defined by the two turning points of the orbit $r_p=p/(1+e)$ and $r_a=p/(1-e)$, where $r_p$ and $r_a$ are respectively the periastron and apastron. The $p$ and $e$ are related to $E$ and $L$ through \cite{Maggiore07}
\begin{align}
p&=\frac{L^2}{GM\mu^2},\label{eq4}\\
e^2&=1+\frac{2EL^2}{G^2M^2\mu^3}\label{eq5}.
\end{align}
The radius $r$ is related to $p$ and $e$ by
\begin{equation}
r=\frac{p}{1+e\cos\phi}\label{eq6}.
\end{equation}

We now introduce the GW back-reaction and the dynamical friction into the IMRI's dynamics which make $E$ and $L$ do not conserved any more. Differentiating the Eqs.~(\ref{eq4}) and (\ref{eq5}) we get
\begin{eqnarray}
\dot{p}=\frac{2L}{GM\mu^2}\dot{L}=2\sqrt{\frac{p}{GM\mu^2}}\dot{L},\label{eq7}\\
\dot{e}=\frac{p}{GM\mu e}\dot{E}+\frac{(e^2-1)}{e\sqrt{GM\mu^2p}}\dot{L}.\label{eq8}
\end{eqnarray}
To calculate $\dot{E}$ and $\dot{L}$, we use the adiabatic approximation which assumes that the radiation and dynamical friction operates at a much longer timescale than the orbital period. We can take $\dot{E}$ and $\dot{L}$ to be the time-averaged rates:
\begin{eqnarray}
\dot{E}=\left\langle\frac{dE}{dt}\right\rangle_{\rm GW}+\left\langle\frac{dE}{dt}\right\rangle_{\rm DF},\label{eq9}\\
\dot{L}=\left\langle\frac{dL}{dt}\right\rangle_{\rm GW}+\left\langle\frac{dL}{dt}\right\rangle_{\rm DF},\label{eq10}
\end{eqnarray}
where the symbol $\langle\rangle$ means the time average and the subscripts $\rm GW$ and $\rm DF$ denote the energy and angular momentum loss due to GW emission and dynamical friction, respectively.

To the leading post-Newtonian order \cite{Maggiore07,PhysRev.136.B1224,PhysRev.131.435}, the GW loss of energy and angular momentum can be expressed as
\begin{align}
&\left\langle\frac{dE}{dt}\right\rangle_{GW}=\nonumber\\
&-\frac{32}{5}\frac{G^4\mu^2M^3}{c^5p^5}(1-e^2)^{3/2}\left(1+\frac{73}{24}e^2+\frac{37}{96}e^4\right),\label{eq11}\\
&\left\langle\frac{dL}{dt}\right\rangle_{GW}=\nonumber\\
&-\frac{32}{5}\frac{G^{7/2}\mu^2M^{5/2}}{c^5p^{7/2}}(1-e^2)^{3/2}\left(1+\frac{7}{8}e^2\right).
\label{eq12}
\end{align}

When the stellar mass object moves through the DM minispike, it gravitationally interacts with DM particles and
this effect is called dynamical friction or gravitational drag \cite{1943ApJ....97..255C}. Because of dynamical friction, the stellar mass object running through the DM halo is decelerated in the direction of its motion and loses its kinetic energy as well as its angular momentum. The dynamical friction force is given by \cite{1943ApJ....97..255C,PhysRevD.91.044045}
\begin{equation}
F_{\rm DF}=\frac{4\pi G^2\mu^2\rho_{DM}(r)\ln\Lambda}{v^2},\label{eq13}
\end{equation}
where $\rm ln\Lambda$ is the Coulomb logarithm, here we choose $\ln\Lambda\simeq10$ \cite{Amaro_Seoane_2007}. $v$ is the velocity of the small BH, which can be obtained from the relation $E=-GM\mu/r+\mu v^2/2$ to get
\begin{align}
v&=\sqrt{\frac{2E}{\mu}+\frac{2G M}{r}}\nonumber\\
&=\sqrt{-\frac{GM(1-e^2)}{p}+\frac{2GM}{p}(1+e\cos\phi)},\label{eq14}
\end{align}
where we have used Eqs.~(\ref{eq4}), (\ref{eq5}) and (\ref{eq6}) in the second step. Based on the above two equations, we can average energy loss rate due to dynamical friction with respect to orbital period
\begin{align}
&\left\langle\frac{dE}{dt}\right\rangle_{\rm DF}=\frac{1}{T}\int_0^T\frac{dE}{dt}|_{DF}dt =\frac{1}{T}\int_0^T F_{\rm DF}v dt\nonumber\\
&=\frac{1}{T}\int_0^T\frac{4\pi G^{3/2}\mu^2\rho_{sp}r_{sp}^\alpha\ln\Lambda(1+e\cos\phi)^\alpha}{p^{\alpha-1/2}M^{1/2}(1+2e\cos\phi+e^2)^{1/2}}dt\nonumber\\
&=(1-e^2)^{3/2}\nonumber\\
&\cdot\int_0^{2\pi}d\phi\frac{2G^{3/2}\mu^2\rho_{sp}r_{sp}^\alpha\ln\Lambda(1+e\cos\phi)^{\alpha-2}}
{p^{\alpha-1/2}M^{1/2}(1+2e\cos\phi+e^2)^{1/2}}.\label{eq15}
\end{align}
In the third step we have used Eqs.~(\ref{eq1}) and (\ref{eq14}). In the last step we have used the relation $\int_0^T\frac{dt}{T}(...)=(1-e^2)^{3/2}\int_0^{2\pi}\frac{d\phi}{2\pi}(1+e\cos\phi)^{-2}(...)$.

Based on $v=\sqrt{\dot{r}^2+r^2\dot{\phi}^2}$ and the geometrical relation we can get the angular momentum loss rate due to dynamical friction $(dL/dt)_{\rm DF}=r\cdot F_{\rm DF}( r\dot{\phi}/v)$. With the same procedure as before, we get
\begin{align}
&\left\langle\frac{dL}{dt}\right\rangle_{DF}=\frac{1}{T}\int_0^T\frac{dL}{dt}|_{DF}dt=(1-e^2)^{3/2}\nonumber\\
&\cdot\int_0^{2\pi}d\phi\frac{2G\mu^2\rho_{sp}r_{sp}^\alpha(1+e\cos\phi)^{\alpha-2}\ln\Lambda}
{p^{\alpha-2}M(e^2+2e\cos\alpha+1)^{3/2}}.\label{eq16}
\end{align}
Substituting Eqs.~(\ref{eq11}), (\ref{eq12}), (\ref{eq15}) and (\ref{eq16}) into Eqs.~(\ref{eq7}) and (\ref{eq8}), we get the dynamical equations for eccentric IMRI under the effect of a DM minispike
\begin{widetext}
\begin{align}
\dot{p}&=-\frac{64}{5}\frac{G^3\mu M^2}{c^5p^3}(1-e^2)^{3/2}\left(1+\frac{7}{8}e^2\right)-\frac{4G^{1/2}\mu\rho_{sp}r_{sp}^\alpha\ln\Lambda}{M^{3/2}p^{\alpha-5/2}}(1-e^2)^{3/2}\int_0^{2\pi} d\phi\frac{(1+e\cos\phi)^{\alpha-2}}{(e^2+2e\cos\phi+1)^{3/2}},
\label{eq17}\\
\dot{e}&=-\frac{304}{15}\frac{G^3\mu M^2}{c^5p^4}(1-e^2)^{3/2}e\left(1+\frac{121}{304}e^2\right)-\frac{4G^{1/2}\mu\rho_{sp}r_{sp}^\alpha\ln\Lambda}{p^{\alpha-3/2}M^{3/2}}(1-e^2)^{3/2}\int_0^{2\pi}d\phi\frac{(e+\cos\phi)(1+e\cos\phi)^{\alpha-2}}{(1+2e\cos\phi+e^2)^{3/2}}.
\label{eq18}
\end{align}
\end{widetext}
In Eqs.~(\ref{eq17}) and (\ref{eq18}), the first terms are due to GW back-reaction and the second terms are from dynamical friction of the DM minispike. Due to the factor $(e+\cos\phi)$, for most $e$ the integration of the second term in Eq.~(\ref{eq18}) is negative at the range of $\alpha$ we used. As a result, the DM minispike tends to increase the eccentricity.

It is also possible to use the semi-major axis $a$ instead of $p$ to describe the dynamics of IMRI. According to the relation $a=p/(1-e^2)$ we can get the dynamical equations for $a$ and $e$:
\begin{widetext}
\begin{align}
\dot{a}&=-\frac{64}{5}\frac{G^3\mu M^2}{c^5 a^3}(1-e^2)^{-7/2}\left(1+\frac{73}{24}e^2+\frac{37}{96}e^4\right)-\frac{4G^{1/2}\mu\rho_{sp}r_{sp}^\alpha \ln\Lambda}{M^{3/2}a^{\alpha-5/2}}(1-e^2)^{2-\alpha}\int_0^{2\pi}d\phi\frac{(1+e\cos\phi)^{\alpha-2}}{(e^2+2e\cos\phi+1)^{1/2}},\label{eq19}\\
\dot{e}&=-\frac{304}{15}\frac{G^3\mu M^2}{c^5a^4}e(1-e^2)^{-5/2}\left(1+\frac{121}{304}e^2\right)-\frac{4G^{1/2}\mu\rho_{sp}r_{sp}^\alpha \ln\Lambda}{M^{3/2}a^{\alpha-3/2}}(1-e^2)^{3-\alpha}\int_0^{2\pi}d\phi\frac{(e+\cos\phi)(1+e\cos\phi)^{\alpha-2}}{(1+2e\cos\phi+e^2)^{3/2}}.\label{eq20}
\end{align}
\end{widetext}

\begin{figure}[htbp]
\centering
\includegraphics[width=0.5\textwidth]{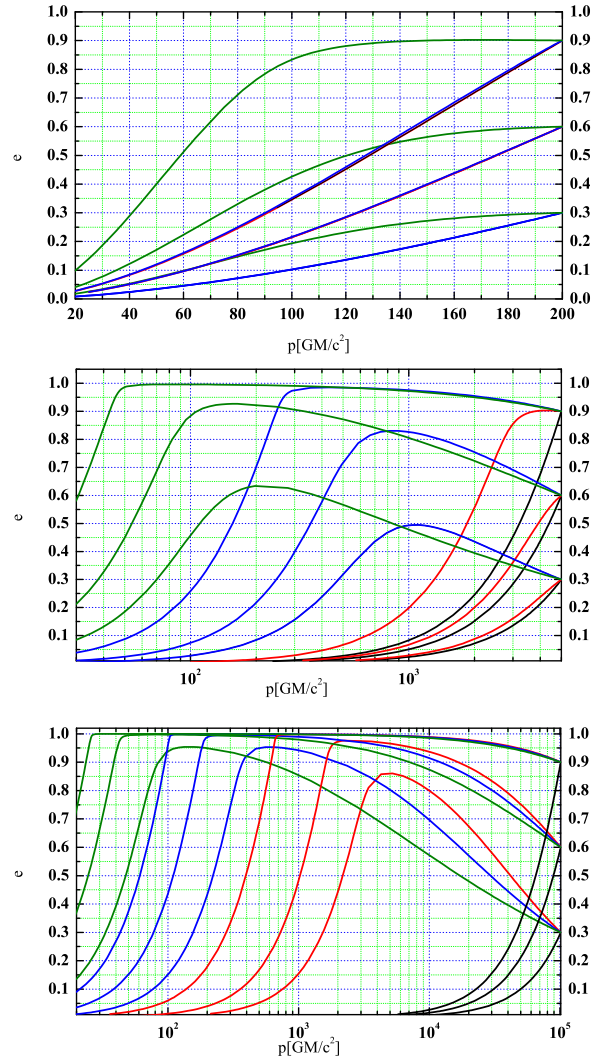}
\caption{The evolution of eccentricity $e$ of an IMRI with different initial $p$ and $e$. The horizontal axis is the semi-latus rectum $p$ with unit of $GM/c^2$ and the vertical axis is the eccentricity. Along time $p$ monotonically decreases, so $p$ can also be looked as a reference time. In this figure we have taken IMBH's mass as $1000M_\odot$ and the small BH's mass $10M_\odot$. The dark lines correspond to cases without DM. The red lines, the blue lines and the green lines correspond to $\alpha=1.5$, $2.0$ and $7/3$ respectively. The upper panel: relatively small initial $p$ is adopted, $p=200 GM/c^2$. The middle panel: intermediate initial $p$ is adopted, $p=5000 GM/c^2$. Lower panel: relatively large initial $p$ is adopted, $p=1\rm AU\simeq 10^5 GM/c^2$.} \label{fig01}
\end{figure}

Fig.~\ref{fig01} depicts the $p-e$ relation for different initial conditions and different profiles of DM minispikes. When the initial $p$ is relatively small, as shown in the upper panel of the Fig.~\ref{fig01}, the curves for  $\alpha=1.5$ and $\alpha=2.0$ are indistinguishable from that without DM. But the relatively denser DM minispike with $\alpha=7/3$ can still reduce the IMRI orbit circularization rate. When the initial $p$ is relatively large as shown in the middle and lower panels of the Fig.~\ref{fig01}, the DM minispike can increase the eccentricity significantly. In some cases the eccentricity can even approach to $1$ along the evolution. When the eccentricity is near $1$, a portion of the three curves overlap together, as shown in the lower panel of the Fig.~\ref{fig01}. If only the initial $p$ is large enough, as an example $1\rm AU$, even the moderate DM density with $\alpha=1.5$ can also increase the eccentricity from $0.3$ to $0.85$.

\begin{figure*}
\centering
\includegraphics[width=0.9\textwidth]{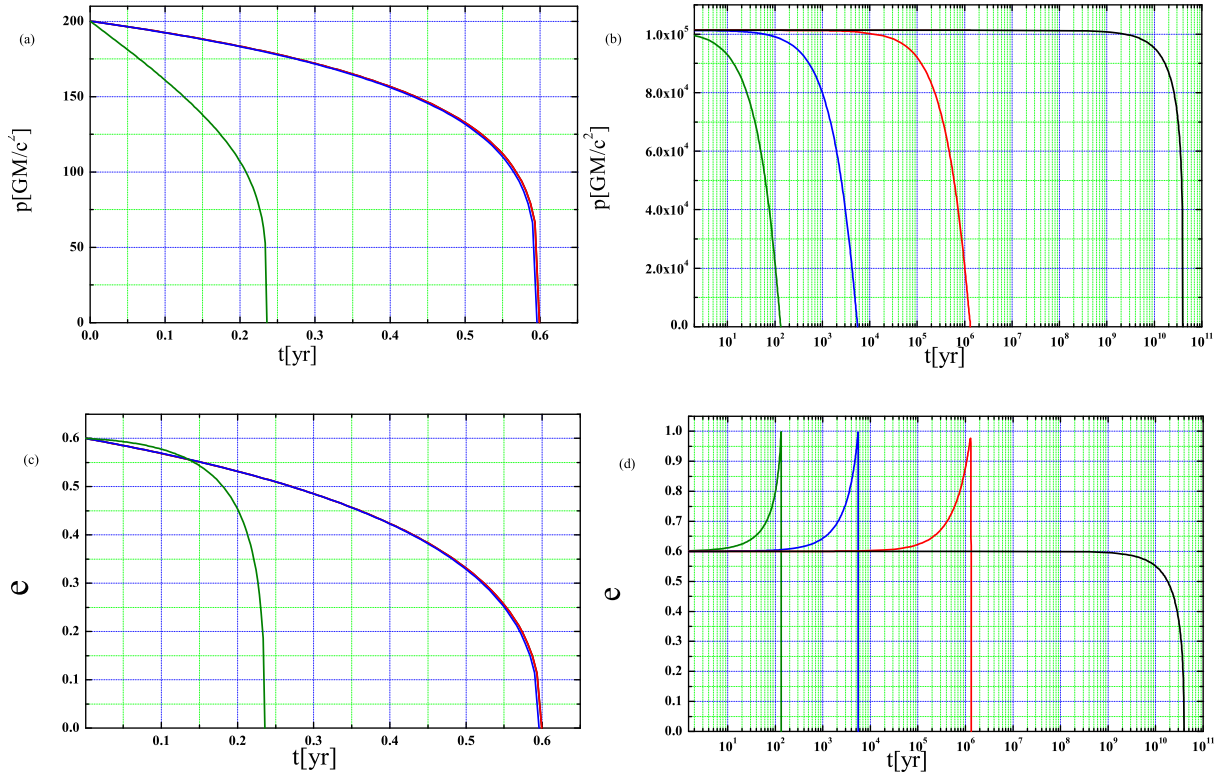}
\caption{The evolution of eccentricity $e$ and semi-latus rectum $p$ of an IMRI with different initial $p$. The horizontal axis is time with unit of $\rm yr$. The vertical axis is semi-rectum $p$ with unit of $GM/c^2$ for upper panels and eccentricity for lower panels. In this figure we have taken IMBH's mass as $1000M_\odot$ and the small BH's mass $10M_\odot$ with initial eccentricity $e=0.6$. The dark lines correspond to cases without DM. The red lines, the blue lines and the green lines correspond to $\alpha=1.5$, $2.0$ and $7/3$ respectively. The left panels: relatively small initial $p$ is adopted, $p=200 GM/c^2$. The right panels: relatively large initial $p$ is adopted, $p=1\rm AU\simeq 10^5 GM/c^2$.} \label{fig02}
\end{figure*}

Fig.~\ref{fig02} indicates the time dependence of $p$ and $e$ for different initial $p$. The left panels show for relatively small initial $p$ only the denser DM minispike with $\alpha=7/3$ influence the evolution obviously. The right panels indicate for relatively large initial $p$, even the moderate DM minispike can expedite the evolution dramatically. With the DM minispike, for most of the time the dynamical friction dominates and the the eccentricity increases. The GW plays a leading role at late times where $p$ is very small and the IMRI merges rapidly.

Interestingly, we find that the dynamical friction can increase the eccentricity. This is because the dynamical friction is inversely proportional to the square of velocity, as shown in Eq.~(\ref{eq13}),so the DM drag is stronger at larger radius. As a result, the energy loss near apastron is larger than that near periastron, and the effect at large radius is dominant. As the loss of energy and velocity near apastron makes the orbit at falling passage steeper than that at rising passage, the eccentricity increases.

It's helpful to compare our results to another similar work investigating the effect of dynamical friction\cite{PhysRevD.98.023536}. In that paper, the authors consider a binary of primordial BHs(mass ratio is $1$) with DM mini-halos attached to each BH as a DM ``dress". In that case, the dynamical friction takes effect only near the periastron where the two BHs pass through the halos of each other. When the two BHs move away from periastron, the loss of velocity drives them to a nearer apastron. As a result, although the friction is small in the close passage, it can still slightly circularize the orbit. We can see the force exerted by DM tends to decrease the eccentricity near periastron and increase it near apastron. The final result depends on the comparison of the effect in the two regions.

The discussions above is also applicable to the GW radiation. As the GW power is larger for shorter distance, it can circularize the orbit. If we take the dynamical friction and GW radiation into consideration together,  whether the eccentricity increases with a given initial condition is determined by the competition of the two mechanisms. The criterion can be obtained by comparing the two terms in Eq.~(\ref{eq18}).  The second term due to friction is proportional to $p^{3/2-\alpha}$ while the first term due to GW is proportional to $p^{-4}$. For a given $\alpha$ we used, when $p$ is relatively large the effect of DM dominates and the eccentricity will increase along time, as shown in the lower panel of Fig.~\ref{fig01} and panel(c) of Fig.~\ref{fig02}. Conversely, the eccentricity decreases, as shown in the upper panel of Fig.~\ref{fig01} and panel(d) of Fig.~\ref{fig02}.

\section{\label{s3}Typical IMRI in a GC}
In Sec.~\ref{s2} we find the DM minispike can increase the eccentricity for large $p$. In this section we investigate a typical IMRI system in a globular cluster (GC).

Recently, there is an increasing number of evidences suggesting that the IMBHs should exist in the centers of GCs or the nuclear stellar clusters (NSCs) of dwarf galaxies \cite{Feng:2011pc,2009Natur.460...73F}. If minispikes exit around these IMBHs, they are less likely to be destroyed by galaxy mergers because their hosts may not have experienced major mergers in the past \cite{PhysRevLett.95.011301,PhysRevD.72.103517}. Motivated by these suggestions, in this section we investigate the evolution of an IMRI in the center of a typical GC and find out the waveform.

Besides the dynamical friction against the DM background and the GW effect, another competing mechanism which could also extract the orbital energy from the IMRI is ``dynamical hardening" \cite{Heggie:1975tg,G_rkan_2006,Mandel:2007hi}. In a GC, after the IMBH form a binary with other stellar objects, for a timescale shorter than the merger timescale an interloper will encounter the binary interacting with them in a complicated way such that by the end of the interaction the interloper is re-ejected into the background, taking away a fraction of the energy and angular momentum from the binary.

We adopt the formula given in \cite{1995ApJ...440..554Q} to calculate the relative efficiency of dynamical hardening:
\begin{equation}
\frac{da}{dt}=-\frac{G H\rho a^2}{\sigma},\label{eq21}
\end{equation}
where $a$ is the semi-major axis of the binary, $H\simeq 15$ is a constant, $\rho$ is the typical density of the background stars and $\sigma$ is the velocity dispersion of these stars. We use the typical parameters in GCs that $\sigma=10\rm km/s$, $\rho=nm_\ast$ where $n=10^{5.5}/\rm pc^3$ is the density of stars and $m_\ast=0.5M_\odot$ is the average mass of a single star \cite{1993ASPC...50.....D}.

When the evolution of the binary is dominated by the hardening process, it shrinks the orbit efficiently. After this process terminates, the binary is driven by GW
reaction and dynamical friction of DM. We can readily separate the evolution of the IMRI according to the two competing processes by comparing their associated timescales: $t_{\rm harden}=|a/\dot{a}_{\rm harden}|$ and $t_{\rm DM}=|a/\dot{a}_{\rm DM}|$, where $\dot{a}_{\rm harden}$ is given by Eq.~(\ref{eq21}) and $\dot{a}_{\rm DM}$ can be found in Eq.~(\ref{eq19}). In \cite{Gultekin_2006} the authors examined the eccentricity of the binary after its final three body interaction and found a typical value of $e\simeq 0.98$. Taking this as the typical value in Eq.~(\ref{eq19}) and equating the two timescales, we can get the threshold of the semi-major axis after the last hardening process, that is $\simeq 3.5\rm AU$ without DM, $\simeq158\rm AU$ for $\alpha=1.5$, $\simeq 1544\rm AU$ for $\alpha=2.0$ and $\simeq 4253 \rm AU$ for $\alpha=7/3$. We can see that denser DM minispikes makes the hardening process of IMRI terminate at larger separation.

To see how the DM minispikes affect the detectability of the gravitational wave signal, we substitute the initial eccentricity of $0.98$ and initial semi-major axis as described above into Eqs.~(\ref{eq19}) and (\ref{eq20}) and integrate until the binary enter the LISA band. For circular orbits, the frequency of gravitational wave emission is mainly twice the orbital frequency, but any harmonics
\begin{align}
&f_{\rm GW}=n\Omega/2\pi,\\
&\Omega=(GM/a^3)^{1/2},\label{eq22}
\end{align}
for eccentric binaries maybe important \cite{Han2017Excitation}. Here $n$ is the harmonic number. $\Omega=2\pi/T$ is the angular frequency of the orbit where $T$ is the orbital period. On the other hand, we have seen that in the history of the evolution of an IMRI the eccentricity could approach to $1$ with appropriate initial conditions. If this does happen, and the large eccentricity can keep until the orbital frequency enters the LISA band, the most powerful harmonic should be $n=1$ as the source will generate bursts of gravitational radiation at each time the object pass through periastron.
\begin{figure}
\centering
\includegraphics[width=0.5\textwidth]{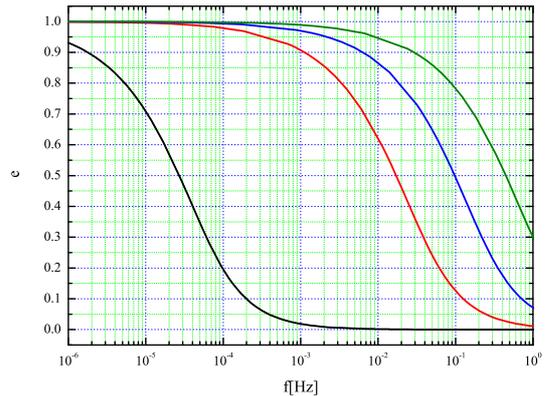}
\caption{The evolution of eccentricity $e$ for a typical IMRI in a GC with different profiles of DM minispike. Here we have adopted IMBH's mass $1000M_\odot$ and the small BH's mass $10M_\odot$. The horizontal axis is the orbital frequency $f$ and the vertical axis is the eccentricity. The dark line is the case without DM with initial conditions $a_0=3.5\rm AU$ and $e_0=0.98$. The red line corresponds to $\alpha=1.5$ with initial conditions $a_0=158\rm AU$ and $e_0=0.98$. The blue line corresponds to $\alpha=2.0$ with initial condition $a_0=1544\rm AU$ and $e_0=0.98$. The green line corresponds to $\alpha=7/3$ with initial conditions $a_0=5253\rm AU$ and $e_0=0.98$. At $10^{-4}\rm Hz$, the lower bound of the LISA band, the eccentricity is respectively $0.185$ without DM, $0.98$ for $\alpha=1.5$, $0.993$ for $\alpha=2.0$ and $0.997$ for $\alpha=7/3$. At $10^{-3}\rm Hz$, the eccentricity is $0.0184$ without DM, $0.907$ for $\alpha=1.5$, $0.97$ for $\alpha=2.0$ and $0.989$ for $\alpha=7/3$.} \label{fig03}
\end{figure}

Considering the above mentioned fact, we obtain the relation between the eccentricity $e$ and the orbital frequency $f$. The Fig.~\ref{fig03} depicts the evolution of the eccentricity $e$ as a function of the orbital frequency $f$ in the range between $10^{-6}\rm Hz$ and $1\rm Hz$. Without DM minispike, when the orbital frequency of the binary reaches the lower side of the LISA band $10^{-4}\rm Hz$, the eccentricity drops to less than $0.2$. When the binary evolves to $10^{-3}\rm Hz$ the eccentricity drops to less than $0.02$. On the contrary, when DM minispike exists, the eccentricity is larger than $0.95$ at $10^{-4}\rm Hz$. At $10^{-3}\rm Hz$, the eccentricity can still keep its value larger than 0.9.

The two independent GW polarization modes in the lowest post-Newtonian order is the function of $r$, $\phi$, $\dot{r}$ and $\dot{\phi}$ \cite{PhysRevD.70.064028}:
\begin{align}
 h_+=&\frac{G \mu}{c^4R}\left\{(1+\cos^2\theta)\left[ \left(\frac{GM}{r}+r^2\dot{\phi}^2-\dot{r}^2\right)\cos(2\phi)\right.\right.\nonumber\\
 &\left.\left.+2\dot{r}r\dot{\phi}\sin(2\phi)\right]-\sin^2\theta\left[\frac{GM}{r}-r^2\dot{\phi}^2-\dot{r}^2\right]\right\}\label{eq23}\\
 h_\times=&2\frac{G\mu\cos\theta}{c^4R}\left\{\left(\frac{GM}{r}+r^2\dot{\phi}^2-\dot{r}^2\right)\sin(2\phi)\right.\nonumber\\
 &\left.-2\dot{r}r\dot{\phi}\cos(2\phi)\right\},\label{eq24}
\end{align}
where $\theta$ denotes the inclination angle of the orbital plane
with respect to the plane of the sky and $R$ is the distance to the source.

When the IMRI moves along an eccentric orbit, the temporal evolution of $r(t)$, $\phi(t)$, $\dot{r}(t)$ and $\dot{\phi}(t)$ is \cite{Maggiore07,PhysRevD.65.084011}
\begin{align}
r&=\left(\frac{GM}{\Omega^2}\right)^{1/3}[1-\cos u],\label{eq25}\\
\phi&=\phi_0+2\arctan\left[\left(\frac{1+e}{1-e}\right)^{1/2}\tan \frac{u}{2}\right],\label{eq26}\\
\dot{r}&=\frac{(GM\Omega)^{1/3}e\sin u}{1-e\sin u},\label{eq27}\\
\dot{\phi}&=\frac{\Omega\sqrt{1-e^2}}{(1-e\sin u)^2},\label{eq28}
\end{align}
where $u$ is called the eccentric anomaly, and is related to $t$ by the famous Kepler equation
\begin{equation}
u-e\sin u=\Omega t.\label{eq29}
\end{equation}
In order to get an explicit expression of $u(t)$, we expand $u$ in terms of $\Omega t$:
\begin{equation}
u=\Omega t+\sum\limits_{s=1}^{\infty} \left(\frac{2}{s}\right)J_s(se)\sin(\Omega t),\label{eq30}
\end{equation}
where $J_s(x)$ is the first kind of Bessel function of order $s$ with $s\geq 1$. With these equations we can get the waveforms for different profiles of DM minispikes. For simplicity, we set the initial polar angular $\phi_0=0$ and the inclination angle $\theta=0$. In Fig.~\ref{fig03} we plot the waveform $h_\times$ of the IMRI with a distance $R=100\rm Mpc$ from us at the orbital frequency $f=10^{-3}\rm Hz$ for different profiles of DM minispikes.

\begin{figure}
\centering
\includegraphics[width=0.5\textwidth]{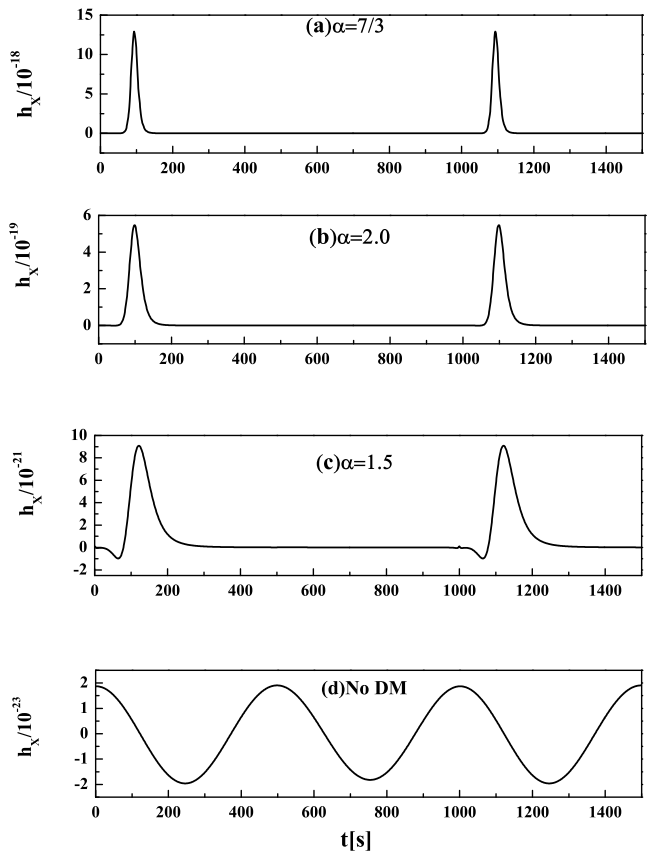}
\caption{The GW waveform $h_\times$ of an IMRI as a function of time $t$ at the orbital frequency of $f=10^{-3}\rm Hz$ corresponds to different profiles of DM minispikes. Again we adopt the IMBH's mass $1000M_\odot$ and the small BH's mass $10M_\odot$. We assume the IMRI locates at the distance $100\rm Mpc$ from us.} \label{fig04}
\end{figure}

As shown in the Fig.~\ref{fig04}, the DM minispike can be easily identified through the detection of the gravitational waveforms. Without DM, the eccentricity is nearly $0$, the waveform is sine wave and the wave frequency is twice of the orbital frequency. When DM minispike exists, the burst of GW emission near periastron generate a cusp in the waveform. The amplitude of the GW waveform is enhanced a lot by the DM minispike. Comparing to the case without DM, the amplitude is increased by two orders even with the moderate DM minispike with $\alpha=1.5$, and in the extremal case the amplitude is increased by $5$ orders. The enhanced amplitude may make the detection of GWs easier.

Finally, it's necessary to point out the feedback of the DM mini-halo may affect the results. This effect is also investigated in the case of primordial BHs\cite{PhysRevD.98.023536}.  Dynamical friction transfers energy from the orbiting BH to the DM particles in the mini-halo, and can in principle heat and unbind part of it. If this effect can substantially reduce the DM minispike, our current results may be changed a lot. However, in the DM minispike the DM particles are much more difficult to be ejected than those around primordial BHs, as they are bound in a much deeper potential well of the IMBH.  On the other hand, energy exchange between stars may regenerate stellar cusps around super massive BHs\cite{Preto_2004}. Whether this can happen to DM particles and replenish DM minispikes needs further checking. The detailed study is beyond the scope of  this paper, but we should still be cautious of this effect.

\section{\label{s4}Summary and conclusions}
In this paper we have considered a system composed of an IMBH surrounded by a DM minispike and a stellar BH orbiting around it. Especially we investigated the evolution of the IMRI's eccentricity. We derived the evolution equations of the IMRI under the effect of dynamical friction against DM minispike under the adiabatic approximation.

The presence of DM minispike tends to increase the eccentricity. The effect is more significant for denser DM minispike and larger semi-rectum $p$. For a sufficiently large initial $p$, even the moderately dense DM minispike can increase the eccentricity dramatically. In the most extremal case an initially moderate eccentricity can be increased to approach to $1$. With quite small initial $p$, the GW back reaction dominates, but the DM minispike can also dramatically reduce the rate of circularization.

We have also studied the evolution of an IMRI in the center of a typical GC. In this case another competing mechanism called dynamical hardening has been taken into consideration. We found the DM minispike can increase the efficiency of extracting orbits, which can make the semi-major axis after the last hardening process much larger.

In this realistic case, the eccentricity of the IMRI can also be increased by the DM minispike and keep its large value until the IMRI evolves into the LISA band. At $10^{-4}\rm Hz$, which is the lower side of LISA band, with DM minispike the eccentricity is larger than $0.95$, and for mildly dense DM minispikes, the value can even come up to $1-e<10^{-2}$. In the contrast, without DM the eccentricity would drop to less than $0.2$.

We also investigated the gravitational waveform for the IMRI in the typical GC at $10^{-3}\rm Hz$, which is the sensitive frequency of LISA. At this frequency the eccentricity is still larger than $0.9$ with DM minispike, and without DM the eccentricity approaches to $0$. The large eccentricity corresponds to the DM minislike cases can produce cusps in the GW waveform and enhance the amplitude by $2$ to $5$ orders, which can be used to distinguish the existence of DM minispike by LISA.

Our calculations show that the DM minispikes may affect the evolution of IMRIs a lot. All the anticipations can be tested by future space-based GW detectors including LISA \cite{lisapage}, Taiji \cite{Gong_2011} and Tianqin \cite{Luo_2016}, which will give a stringent constraint to the physical models of DM.

\acknowledgments
This work was supported by the NSFC (No.~11690023 and No.~11622546). Z Cao was supported by ``the Fundamental Research Funds for the Central Universities".


\end{document}